\documentclass[12pt,preprint]{aastex}

\usepackage{amsmath,amssymb}
\usepackage{graphics}
\shorttitle{Electron Spectrum in 3C 279}
\shortauthors{Kusunose, Takahara, and Kato}

\begin{document}

\title{The Electron Spectrum in 3C 279 and the Observed Emission Spectrum}

\author{Masaaki Kusunose\altaffilmark{1,2}, 
Fumio Takahara\altaffilmark{3},
and Tomohiro Kato\altaffilmark{1,4}}

\altaffiltext{1}{Department of Physics, School of Science and Technology,
Kwansei Gakuin University, Sanda 669-1337, Japan}
\altaffiltext{2}{kusunose@kwansei.ac.jp}
\altaffiltext{3}{Department of Earth and Space Science, 
Graduate School of Science, Osaka University,
Toyonaka 560-0043, Japan; takahara@vega.ess.esi.osaka-u.ac.jp}
\altaffiltext{4}{scms1009@ksc.kwansei.ac.jp}

\begin{abstract}
The emission mechanisms of the blazar 3C 279 are studied
by solving the kinetic equations of electrons and photons 
in a relativistically moving blob.  
The $\gamma$-ray spectral energy distribution (SED) is fitted 
by inverse Compton scattering of external photons.  
The bulk Lorentz factor of the emitting blob is found to be $\sim 25$, and 
the magnetic field is found to be $\sim 0.3$ G.  
GeV $\gamma$-rays are well explained by 
inefficiently cooled electrons because of the Klein-Nishina effects. 
The electron spectrum is not a broken power law with a steeper spectrum 
above a break energy, which is often used to fit the observed SED.  
The kinetic energy density of the nonthermal electrons dominates 
the magnetic energy density; this result is qualitatively the same as that 
for TeV blazars such as Mrk 421 and Mrk 501.  
The $\gamma$-ray luminosity of 3C 279 is often observed to increase rapidly.  
We show that one of the better sampled $\gamma$-ray flares can be explained
by the internal shock model.
\end{abstract}


\section{Introduction}

Blazars are active galactic nuclei characterized by
high energy emission and short time variability
\citep[e.g.,][]{krolik99}.
Recent observations of blazars show that they are powerful sources
of high energy emission \citep[][for review]{umu97},
which is explained by nonthermal emission by particles in relativistic jets.
The emission of blazars in the GeV energy range was observed by EGRET 
onboard the {\it Compton Gamma Ray Observatory} \citep{thom95,muk97,har01}.
These observations show that a large fraction of emission power from
blazars is in $\gamma$-ray band 
(see Ghisellini et al. 1998 and Collmar 2001 for reviews).

Multiwavelength observations of 3C 279 (redshift $z = 0.538$) were performed 
several times \citep{mar94,wer98,har01}. 
Based on these results, $\gamma$-rays from 3C 279 are found to be explained 
by inverse Compton scattering of external UV radiation \citep[e.g.,][]{it96}.
The sources of seed soft photons (e.g.,
accretion disks \citep{ds93},
broad-line regions \citep{sbr94},
and beamed radiation reflected by broad-line region clouds \citep{gm96})
have been proposed by many authors.
Recently, \cite{har01} assumed soft photons from both 
the accretion disk and the broad-line region in order to 
model the emission spectra from 3C 279.
\cite{bal02}, on the other hand, applied a broad-line region model
to 3C 279.

Most emission models have assumed that electrons obey a power law 
\citep[e.g.,][]{har01} or a broken power law \citep[e.g.,][]{it96,bal02}.
However, an alternative electron spectrum is possible as a result of
the Klein-Nishina (K-N) effects;
the smaller cooling rate due to the K-N effects makes the electron 
spectrum harder \citep{da02}, 
and the emissivity of scattered photons is smaller \citep[][]{gkm01}.
Thus, the electron spectrum must be solved self-consistently,
including the cooling rate in the K-N regime.

In previous work, we solved the kinetic equations of electrons 
and photons simultaneously 
in a relativistically moving blob (Li \& Kusunose 2000;
Kino, Takahara, \& Kusunose 2002, hereafter KTK).
Here we include the inverse Compton scattering of external soft photons
(external Compton scattering, hereafter EC)
and obtain the physical quantities for 3C 279 such as the bulk Lorentz factor,
the energy densities of electrons and magnetic fields, etc.

The variability of blazars on small timescales is often observed 
\citep[e.g.,][]{wer98}.
The time variation of fluxes is not only useful for obtaining constraints
on the size of the emission region but is also important for studying
the emission and particle acceleration mechanisms 
\citep{mk97,kus00,lk00,bc02}.  Flares observed in X-ray and $\gamma$-ray
regimes might be explained by the internal collisions of plasma blobs in a jet
and the subsequent cooling.  Without considering the details of 
particle acceleration after the collision, it is possible to
obtain the emission spectrum by assuming that the time evolution occurs
quasi-steadily.  In this Letter we present a model to explain the flare
of 3C 279 observed in 1996 February.

In \S 2, kinetic equations for electrons and photons are described.
Numerical results are given in \S3 for a steady state.
Results for time variability are shown in \S4.
Finally, \S5 is devoted to the summary of our results and discussion.

%%%%%%%%%%%%%%%%%%%%%%%%%%%%%%%%%%%%%%%%%%%%%%%%%%%%%%%%%%%%%%%%%%%%%

\section{Model} \label{sec:model}

The formulation used in this Letter is the same as in \cite{lk00}
and KTK.
We assume that a spherical blob with radius $R$ 
moves with a relativistic speed with Lorentz factor $\Gamma$.
Accelerated electrons obeying a power law are injected 
into the blob uniformly, and they are cooled by synchrotron emission
and inverse Compton scattering.

\subsection{Kinetic Equations} \label{sec:kinetic}

The equation describing the time evolution of the electron number 
spectrum in the blob is given by 
\begin{equation}
\label{eq:elkinetic}
\frac{\partial n_e(\gamma)}{\partial t}
= - \frac{\partial}{\partial \gamma}
 \left[ 
\left( \frac{d\gamma}{dt} \right)_{\rm loss}  n_e(\gamma) \right]
- \frac{n_e(\gamma)}{t_{e, {\rm esc}}} + q(\gamma) \, , 
\end{equation}
where $\gamma$ is the Lorentz factor of electrons and $n_e(\gamma)$ is 
the number density of electrons per unit $\gamma$. 
We assume that nonthermal electrons with a power law
are injected at a rate:
\begin{equation}
q(\gamma) = q_0 \, \gamma^{-p} \exp (- \gamma/\gamma_\mathrm{max})
\qquad \text{for} \,\,\, \gamma \ge \gamma_\mathrm{min} \, ,
\end{equation}
where $q_0$ is the normalization.
The injection is assumed to continue during a simulation 
with the rate given above.
The energy-loss rate of electrons is denoted by $(d \gamma / dt)_{\rm loss}$,
which is due to synchrotron radiation and inverse Compton scattering. 
The synchrotron-loss rate is calculated by the formulation given by
\cite{rm84} for mildly relativistic electrons and by \cite{cs86} for
relativistic electrons.
The cooling rate of inverse Compton scattering is calculated by \cite{cb90};
although the EC process is anisotropic, we use the formulation for 
isotropic scattering.
The value of the electron-escape timescale, $t_{e, \mathrm{esc}}$, 
is set to be $3 R/c$, assuming that the escape of electrons from the jet is 
likely due to advection.  
We use a method developed by \cite{xiao96} for numerical calculations.

To obtain the isotropic photon field in the blob frame, i.e., except
the EC component, we solve the following equation:
\begin{equation}
\label{eq:phkinetic}
\frac{\partial n_{\rm ph}(\varepsilon)}{\partial t} = \dot{n}_{\rm C}(\varepsilon)
+ \dot{n}_{\rm em}(\varepsilon) - \dot{n}_{\rm abs}(\varepsilon) 
- \frac{n_{\rm ph}(\varepsilon)}{t_{\gamma, {\rm esc}}} \, , 
\end{equation}
where $n_{\rm ph}(\varepsilon)$ is the photon number density per 
unit energy $\varepsilon$; 
Compton scattering and synchrotron emission are
denoted by $\dot{n}_{\rm C}(\varepsilon)$ and $\dot{n}_{\rm em}(\varepsilon)$,
respectively; $\dot{n}_{\rm abs}(\varepsilon)$ is absorption
due to the self-absorption of synchrotron process and pair production;
and $n_{\rm ph}(\varepsilon)/t_{\gamma, {\rm esc}}$ denotes 
the escape of photons (we set $t_{\gamma, {\rm esc}} = R/c$).
The EC component is separately
calculated following Georganopoulos et al. (2001; see also
Dermer \&  Schlickeiser 2002)
because the external photons are highly anisotropic in the blob frame.
We assume that in the rest frame of the central black hole, 
the external radiation is isotropic with energy density 
$u_\mathrm{ext}$ and monochromatic with energy $\varepsilon_\mathrm{ext}$.
In the blob frame, the comoving energy density of the soft photons is
$u^\prime_\mathrm{ext} = u_\mathrm{ext} \Gamma^2 (1 + \beta_\Gamma^2/3)$,
where $\beta_\Gamma = (1-\Gamma^{-2})^{1/2}$ \citep{ds02}.
On the other hand, the energy of the external photons is
$\varepsilon^\prime_\mathrm{ext} \approx \Gamma \varepsilon_\mathrm{ext}$.

The observed radiation is boosted by the Doppler effect characterized
by the beaming factor 
${\cal D} = [\Gamma (1 - \beta_\Gamma \cos \theta_\mathrm{obs})]^{-1}$,
where $\theta_\mathrm{obs}$ is
the angle between the jet propagation and the line of sight \citep{bk79}.
In this Letter, we assume ${\cal D} = \Gamma$.
The parameters to be determined are
$\Gamma$, $\gamma_\mathrm{min}$, $\gamma_\mathrm{max}$, $p$, $q_0$, $B$, $R$,
$\varepsilon_\mathrm{ext}$, and $u_\mathrm{ext}$.

The comoving quantities are transformed back into the observer's frame
depending on the beaming factor and the redshift.
The observed photon energy
$\varepsilon_{\rm obs} = \varepsilon \, {\cal D}/(1+z)$, 
and the time duration $dt_{\rm obs} = dt \, (1+z)/{\cal D}$.

%%%%%%%%%%%%%%%%%%%%%%%%%%%%%%%%%%%%%%%%%

\section{Numerical Model of a Steady State} \label{sec:preflarefit}

We use the observational data in \cite{har01}.
In particular, the spectral energy distribution (SED) 
observed between 1996 January 16 and 30
(P5a in their paper) is used for our model of a steady state.
3C 279 is in a preflare stage in this period
and is not necessarily in a quiescent sate.
However, we fitted the data assuming that the spectrum is approximated 
by a steady state solution of the kinetic equations.

\begin{figure}
\includegraphics[scale=0.7]{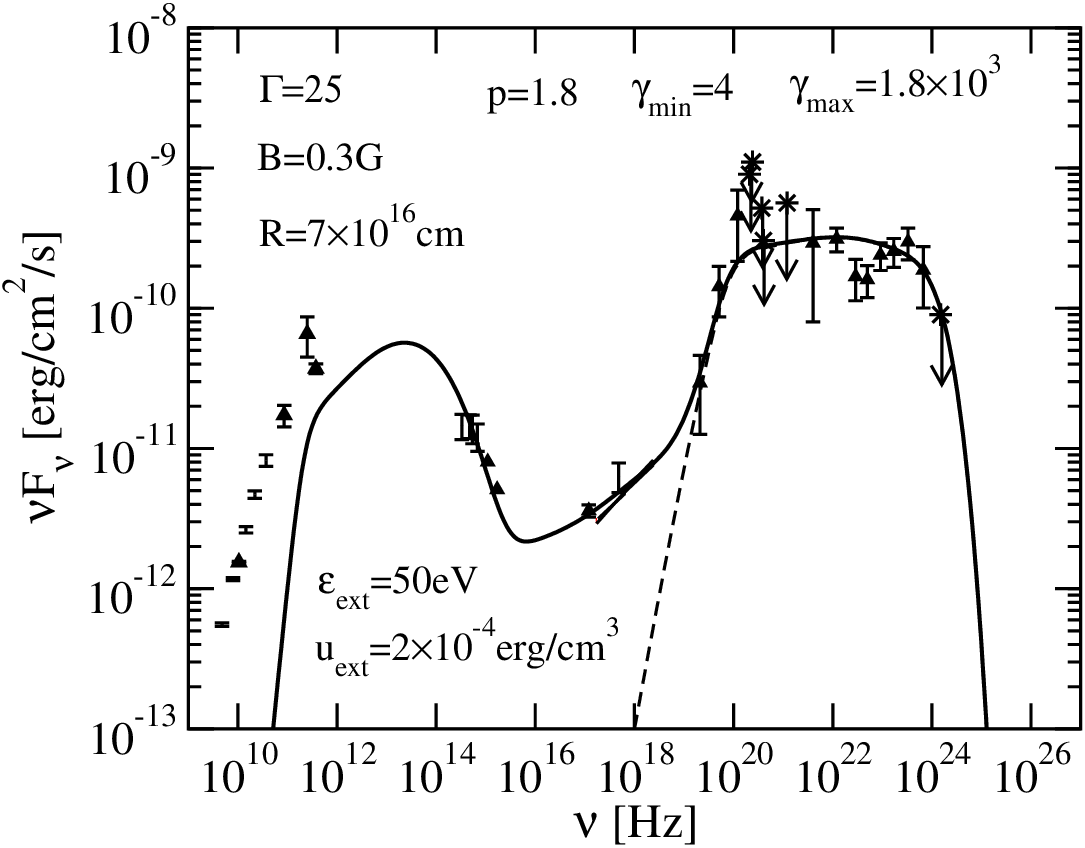}
\caption{SED observed between 1996 January 16 and 30 
with our model ({\it solid line}):
The data are from \cite{har01}. 
The values of the parameters are shown in the figure.
The spectral component of EC scattering is shown by the dashed line.
\label{fig:sed-p5a}}
\end{figure}

In Figure \ref{fig:sed-p5a}, our model is plotted with the observed data.
We obtain the following values for the parameters:
$\Gamma = 25$, $B = 0.3$ G, $R = 7 \times 10^{16}$ cm, $p = 1.8$, 
$\gamma_\mathrm{min} = 4$, $\gamma_\mathrm{max} = 1.8 \times 10^3$, 
the injection rate $2.3 \times 10^{-4} \, \text{cm}^{-3} \text{s}^{-1}$, 
$\varepsilon_\mathrm{ext} = 50$ eV, and
$u_\mathrm{ext} = 2 \times 10^{-4} \, \text{ergs} \,\,\text{cm}^{-3}$.
The spectrum below $\sim 10^{15}$ Hz is produced by synchrotron radiation,
although the radio emission is probably from the outer regions of the jet.
The radiation between $10^{16}$ and $10^{19}$ Hz 
is from synchrotron-self-Compton scattering, 
and the $\gamma$-rays above $10^{20}$ Hz are from the inverse Compton
scattering of external soft photons.
As shown in Figure \ref{fig:sed-p5a} by the dashed line,
the photons scattered by electrons with $\sim \gamma_\mathrm{min}$
appear around $10^{20}$ Hz.  Thus the flux around $10^{20}$ Hz is 
sensitive to the value of $\gamma_\mathrm{min}$.
However, because the data in this regime are mostly upper limits,
the constraint on $\gamma_\mathrm{min}$ is weak.

Based on our numerical calculations, it is found that
$u_\mathrm{kin}/u_{B} = 36$, 
where $u_\mathrm{kin}$ is the energy density of nonthermal electrons
and $u_{B}$ is the magnetic energy density.
We also find that the Poynting power is given by 
$L_\mathrm{Poy} = \pi R^2 c u_{B} {\cal D}^2 
= 1.03 \times 10^{45} \, \text{ergs} \,\, \text{s}^{-1}$ 
and that the kinetic power is given by
$L_\mathrm{kin} = \pi R^2 c u_\mathrm{kin} {\cal D}^2 
= 3.69 \times 10^{46} \, \text{ergs} \,\, \text{s}^{-1}$.
The dominance of the kinetic energy of nonthermal electrons over 
that of magnetic fields is the same as that for TeV blazars such as Mrk 421 
and Mrk 501 (KTK).

In Figure \ref{fig:el-p5a}, the electron spectrum for the SED shown in 
Figure \ref{fig:sed-p5a} is presented.
It is found that the electron spectrum is rather flat 
in the $\gamma$-$\gamma^2 n_e(\gamma) $ plot.
This is because the cooling of high-energy electrons is inefficient, 
owing to the K-N effects \citep{blu71}.
Note that the external photon energy in the blob frame
is $\varepsilon^\prime \sim \Gamma \varepsilon_\mathrm{ext}$ and
that the K-N effects become effective for electrons with
$\gamma_\mathrm{K-N} \sim m_e c^2 /\varepsilon^\prime \sim 400$,
which is less than $\gamma_\mathrm{max}$.
The spectral shape is contrary to the models often employed 
to fit the observed data of blazars, 
in which $n_e(\gamma) \propto \gamma^{-p}$ 
below a break energy $\gamma_\mathrm{br}$ 
and $n_e(\gamma) \propto \gamma^{-(p+1)}$ 
above $\gamma_\mathrm{br}$;
the value of $\gamma_\mathrm{br}$ is determined by 
$t_\mathrm{cool} = t_{e, \mathrm{esc}}$, 
where $t_\mathrm{cool}$ is the electron cooling time.
Nominally, $\gamma_\mathrm{br}$ is about 30.
As seen in Figure \ref{fig:el-p5a}, the actual spectral shape is much
different from this conventional expectation.
The flat $\gamma$-ray spectrum in $10^{21} - 10^{24}$ Hz in the $\nu$-$F_\nu$ 
plot is the result of these K-N effects. 

The optical depth for photon absorption against electron-positron 
pair production is calculated \citep[][]{gs67}.
The optical depth is much smaller than unity and $\gamma$-rays escape without
absorption.

\section{Flare} \label{sec:flare}

The time variability of emission spectra from blazars is quite common.
3C 279 also shows time variation from radio to $\gamma$-rays
\citep[e.g.,][]{wer98}.
To understand the jet formation mechanisms and particle acceleration processes,
it is important to understand the mechanism that causes the time variability.
\cite{sb01} and \cite{sglc01}
show how flare light curves behave according to the shock-in-jet models, 
which assume that blobs with different speeds collide and
that the kinetic energy of the bulk motion is dissipated.
\cite{bal02} interpreted the time variability of 3C 279
as being the result of various values of $\Gamma$, although they assumed 
a broken power-law spectrum of electrons
without solving the kinetic equation.

\begin{figure}
\includegraphics[scale=0.7]{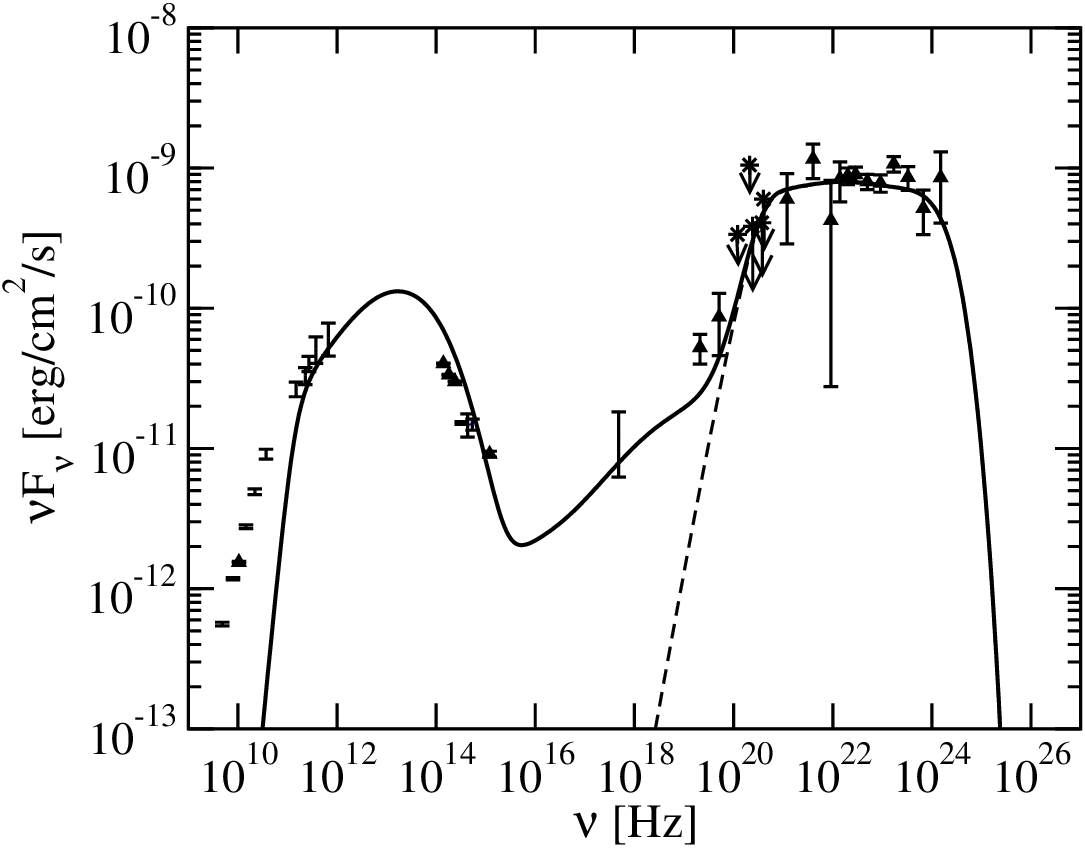}
\caption{SED of the flare between 1996 January 30 and February 6.
The solid line is calculated assuming $\Gamma_f/\Gamma_q = 1.6$.
The spectral component of 
EC scattering is shown by the dashed line.
\label{fig:sed-p5b}}
\end{figure}

We apply the internal shock model, by which \cite{taka03} will explain
the time variability of Mrk 421, to a flare observed from 3C 279.
Without considering the details of the collision process,
we simply assume that a coalesced blob has a bulk Lorentz factor $\Gamma_f$
and that various physical quantities scale with $\Gamma$.
We examine whether or not the flares are explained 
by a change in $\Gamma$, keeping a constant opening angle for the jet cone.  
The distance of the collision location from the base of the jet, $d$, 
scales proportional to $\Gamma^2$,
while the lateral size of the observed region behaves as $R \propto \Gamma$,
which is roughly equal to the shell thickness in the comoving frame,
because the angle from the line of sight is limited by 
$\sim \Gamma^{-1}$.  The soft photon energy density changes as 
$u_\mathrm{ext} \propto \Gamma^{2} d^{-2} \propto \Gamma^{-2}$.
We assume that the kinetic and Poynting powers scale
proportional to $\Gamma^2$ 
so that the total number flux remains constant.  
Thus, the injection rate in the comoving frame, $q$, 
scales proportional to $d^{-2} R^{-1} \propto \Gamma^{-5}$, 
and the magnetic field scales proportional to $\Gamma^{-2}$.
We also assume $\gamma_\mathrm{min}$ and $\gamma_\mathrm{max}$
do not depend on $\Gamma$.
The value of $\varepsilon^\prime_\mathrm{ext}$ scales proportional to
$\Gamma$ because of the Doppler effects.
Using this scaling, we calculate a flare spectrum.
Assuming that the spectrum shown in Figure \ref{fig:sed-p5a} is
in a quiescent state, the spectrum observed between 1996 
January 30 and February 6 is fitted.
Here the bulk Lorentz factor in Figure \ref{fig:sed-p5a} is denoted 
as $\Gamma_q$, and that of a shocked blob is given by $\Gamma_f$.
When $\Gamma_f = 1.6 \Gamma_q = 40$,
the $\gamma$-ray spectrum is fitted fairly well by our model
(Figure \ref{fig:sed-p5b}), although there are some minor discrepancies:
The slope in the optical-UV regime is slightly different;
the flux between $10^{19}$ and $10^{20}$ Hz is underestimated;
and the flux between $10^{20}$ and $10^{21}$ Hz is overestimated.

%%%%%%%%%%%%%%%%%%%%%%%%%%%%%%%%%%%%%%%%%%%%%%%%%%%%%%%%%%%%%%%
\section{Summary and Discussion} \label{sec:concl}

We used the kinetic equations of photons and electrons to calculate
emission spectra from 3C 279, assuming that the emission is from 
a relativistically moving blob almost along the line of sight.
Our numerical solution shows that the $\gamma$-rays are produced by
inverse Compton scattering of external soft photons, 
which is consistent with previous work \cite[e.g.,][]{it96,har01,bal02}.
However, we found that the electron spectrum is not a broken power law
with a steeper spectrum above a break energy.
Although electrons are cooled before escape  
(except those in the lowest energy range), 
the cooling efficiency for high-energy electrons
decreases because of the K-N effects.
As a result, a flat GeV $\gamma$-ray spectrum is obtained 
in the $\nu$-$\nu F_\nu$ plot.
It should be noted that there is an alternative model of GeV emission
that assumes different soft photon sources, i.e.,
the accretion disk and broad line regions \citep{har01}.
In their model, because of the different temperatures of soft photons, 
a broad GeV emission is formed by inverse Compton scattering.

Our result of $\varepsilon_\mathrm{ext} = 50$ eV seems to be a little high
compared with the conventional value of around 10 eV.
However, the spectra of ionizing radiation that form the broad line clouds are
not directly constrained observationally but should extend to energy high
enough to multiply ionize atoms of heavy elements.  From a theoretical point 
of view, even the emission from the accretion disk is known to deviate from 
the blackbody because of scattering effects.  Thus, our choice may be suitable.
Recently, it was suggested that IR radiation from dust might be important,
if the energy dissipation occurs mainly far away from the central region
\citep{bsmm00,sbmm02}; although we have not examined this case, the K-N effects
will be weaker for this component.

The kinetic energy density of nonthermal electrons in the blob is 
an order of magnitude larger than 
the magnetic energy density, which is the same result as for TeV
blazars (KTK).  As a result, the particles transfer more energy
in the jet from the central region to the outer region than 
the magnetic fields do.

We demonstrated that the flare observed between 1996
January 30 and February 6
is explained fairly well by the internal shock model with simple scaling laws
described by a change in the bulk Lorentz factor.
It is remarkable that the flare is fitted with those simple assumptions.

\begin{figure}
\includegraphics[scale=0.7]{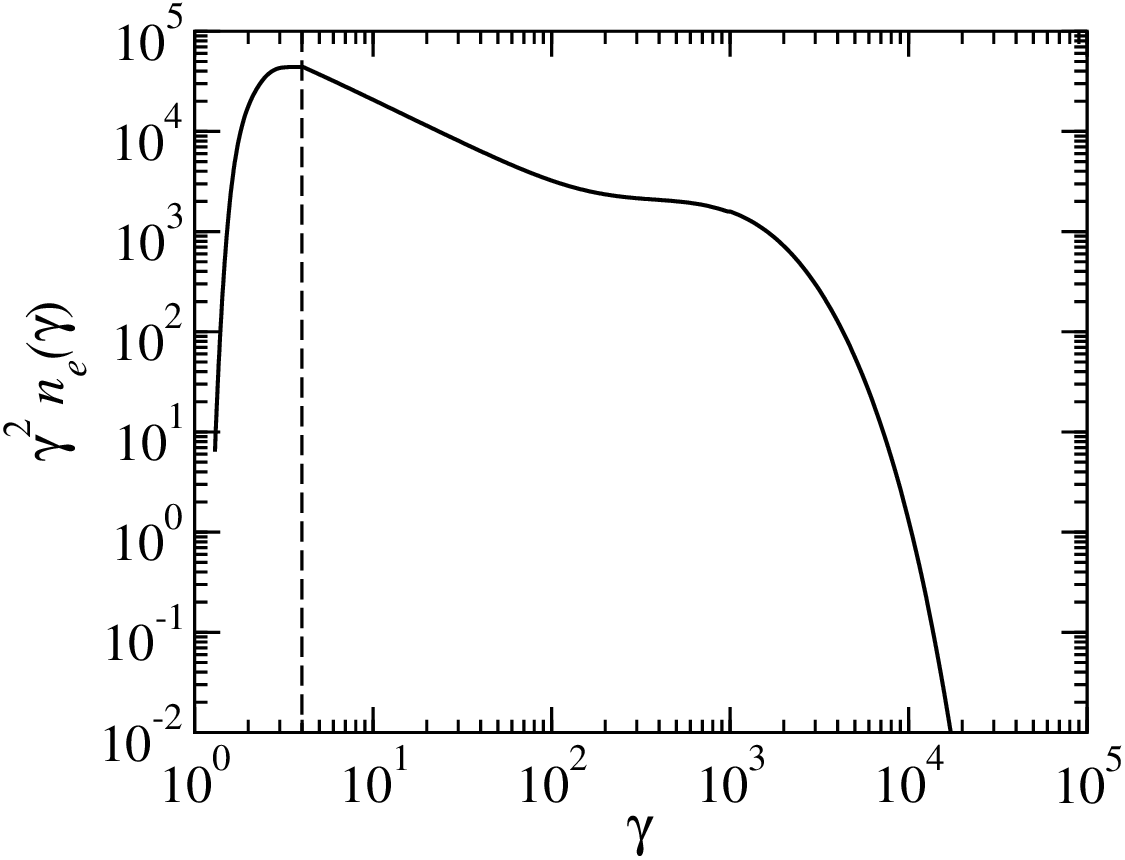}
\caption{Electron spectrum for the SED shown in Figure \ref{fig:sed-p5a}.
The vertical dashed line shows $\gamma_\mathrm{min}$.
The spectrum is flatter above $\gamma \sim 100$ because of the K-N effects.
\label{fig:el-p5a}}
\end{figure}

It was pointed out by \cite{sm00} that the bulk motion of
cold electrons/positrons in the jet scatters the external photons, 
resulting in observed emission peaking at energy 
$\varepsilon_\mathrm{BC} \sim \Gamma^2 \varepsilon_\mathrm{ext}$.
Our model with $\Gamma = 25$ and $\varepsilon_\mathrm{ext} = 50$ eV implies
$\varepsilon_\mathrm{BC} \sim 31$ keV 
($\nu_\mathrm{BC} \sim 7.5 \times 10^{18}$ Hz).
The observed luminosity of bulk Compton (BC) scattering is estimated as
\begin{equation}
\begin{split}
L_\mathrm{BC} & \approx \Gamma^2 \int 
\left( \frac{4}{3} \, c \, \sigma_\mathrm{T} \,
u_\mathrm{ext} \, \Gamma^2 \right) n_c \, dV \\
& \approx 3.0 \times 10^{39} 
\left(\frac{\Gamma}{25}\right)^4 
\frac{u_\mathrm{ext}}{2 \times 10^{-4} \, \text{ergs} \, \text{cm}^{-3}}
\left(\frac{R}{7 \times 10^{16} \text{cm}}\right)^3 \, 
n_{c} \, \,
\text{ergs} \, \text{s}^{-1} \, ,
\end{split}
\end{equation}
where $\sigma_\mathrm{T}$ is the Thomson cross section, 
$n_{c}$ is the number 
density of cold electrons/positrons,
and the integration is done over the blob volume.
For 3C 279 with luminosity distance $d_L = 2.4 \times 10^3$ Mpc, 
where the Hubble constant
$H_0 = 75 \, \text{km} \, \text{s}^{-1} \, \text{Mpc}^{-1}$ is assumed,
$\nu F_\nu$ at $\varepsilon_\mathrm{BC}$ is 
$\sim L_\mathrm{BC}/(4 \pi d_L^2) \sim 7 \times 10^{-14}
n_c/n_e$ $\text{ergs } \, \text{cm}^{-2} \, \text{s}^{-1}$,
where $n_e = 1.65 \times 10^{4} \, \text{cm}^{-3}$ is the
number density of nonthermal electrons shown in Figure \ref{fig:el-p5a}.
The value of $n_c/n_e$ is unknown, 
and unless it is about 100 or more, the BC spectrum is not 
observable; note that the observed value 
is $\sim 10^{-11} \text{ergs} \, \text{cm}^{-2} \, \text{s}^{-1}$
at $\varepsilon_\mathrm{BC} \sim 31$ keV.
Regions between the base of the jet and the blob may contribute to BC
scattering
because more cold electrons may exist there than in the blob.
Let us assume that the bulk Lorentz factor is constant and that 
the number density of the cold electrons 
behaves as  $n_c \propto \ell^{-2}$ for $\ell > \ell_0$, 
where $\ell$ is the distance from the central region 
and $\ell_0$ is the critical distance where the bulk Lorentz factor saturates 
or the jet becomes optically thin. 
If we assume that $u_\mathrm{ext}$ scales proportional to $\ell^{-2}$, 
the predicted value of $L_{\rm BC}$ is $d/\ell_0$ times larger than that 
of equation (4), noting that $n_c dV$ is measured in the comoving 
frame of the blob.
Even if there are no cold electrons is the blob, 
in the inner-jet regions there should exist a corresponding 
number of cold electrons to relativistic electrons in the blob. 
Thus, we obtain the constraint of $\ell_0 >0.01 d$. 
If $n_c/n_e$ in the blob is more than 100, 
too much BC emission is predicted, which probably means that $u_{\rm ext}$ 
does not increase much with a decrease in $\ell$. 
In this case some modifications of our model are needed.

%%%%%%%%%%%%%%%%%%%%%%%%%%%%%%%%%%%%%%%%%%%%%%%%%%%%%%%

\acknowledgements

This work has been partially supported by Scientific Research Grants 
(M.K. and F.T.: 13440061; M.K.:15037210; F.T.: 14079205 and 14340066) from 
the Ministry of Education, Culture, Sports, Science and Technology of Japan.

%%%%%%%%%%%%%%%%%%%%%%%%%%%%%%%%%%%%%%%%%%%%%%%%%%%%%%%

% Figure 1

% Figure 2

% Figure 3

\end{document}